\documentclass[11pt]{cernrep}
\usepackage{amsmath,amssymb,citesort,enumerate,here}
\usepackage{graphicx,wrapfig}
\usepackage{hhline,multirow}

\begin{document}


\title{CRONIN EFFECT IN PROTON-NUCLEUS COLLISIONS: \\
A SURVEY OF THEORETICAL MODELS\footnote{\ \ 
Contribution to the CERN Yellow report on Hard Probes in Heavy Ion Collisions
at the LHC.}
}

\author{A. Accardi}

\institute{Institut f\"ur Theoretische Physik der Universit\"at
Heidelberg,\\ Philosophenweg 19, D-69120 Heidelberg, Germany.\\
e-mail: accardi@tphys.uni-heidelberg.de} 

\maketitle

\begin{abstract}
I compare the available theoretical models that describe the Cronin
effect on hadron and minijet production in proton-nucleus collisions,
pointing out similarities and differences among them. The
effect may be summarized by the value of two variables. Their values
computed in the different models are compared in the energy range 27.4
GeV - 5500 GeV. Finally, I propose to use the pseudorapidity
systematics as a further handle to distinguish among the models.  
\end{abstract}


\section{INTRODUCTION}

\begin{wrapfigure}{r}{5.3cm}
\begin{center}
\vskip-.4cm
\includegraphics[width=5cm]{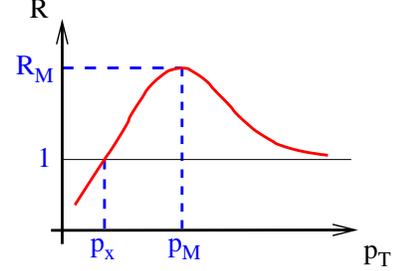} 
\vskip-.3cm
\caption{Definition of $p_\times, p_M,
R_M$.}
 \label{fig:cronineffect}
\end{center}
\vskip-.7cm
\end{wrapfigure}

In this short note I will compare available theoretical models for the
description of the so-called Cronin effect \cite{Cronin75} 
in inclusive hadron spectra
in proton-nucleus ($pA$) collisions. The analysis will be limited to
references containing quantitative predictions in the case of $pA$
collisions \cite{Wang00,ZFPBL02,KNST02,AT01,GV02}. The observable which I
am interested in is the {\it Cronin ratio}, $R$, of the inclusive
differential cross-sections for proton scattering on two different 
targets, normalized to the respective atomic numbers $A$ and $B$:
\begin{equation*}
    R(p_T) = \frac{B}{A} 
        \frac{d\sigma_{pA}/d^2p_T}{d\sigma_{pB}/d^2p_T} \ .
\end{equation*}
In absence of nuclear effects one would expect $R(p_T)$$=$$1$, 
but for A$>$B a suppression is observed experimentally at small $p_T$, 
and an enhancement at moderate $p_T$ with $R(p_T) \rightarrow 1$ 
as $p_T\rightarrow\infty$.
This behaviour may be charachterized by the value of three
parameters: the transverse momentum $p_\times$ at which $R$ crosses unity and the transverse momentum $p_M$
at which $R$ reaches its maximum value $R_M$$=$$R(p_M)$, see
Fig.~\ref{fig:cronineffect}. These {\it Cronin parameters} will be 
studied in Sec.~\ref{sec:predictions}

The Cronin effect has received renewed interest after the 
experimental discovery at RHIC of an unexpectedly small $R<1$ 
in Au-Au collisions at $\sqrt s $$=$$ 130$ GeV compared to $pp$ collisions at
the same energy \cite{SupprCronin}. This fact has been
proposed as an experimental signature of a large jet quenching
suggesting that a Quark-Gluon Plasma was created during the collision
\cite{VGL02}. 
However, the extrapolation to RHIC and LHC energies of the known Cronin effect
at lower energies is haunted by large theoretical uncertainties,
which may make unreliable any interpretation of signals of this kind.
Some light will be shed on this problem by the upcoming RHIC data on
$dA$ collisions at $\sqrt s $$=$$ 200$ GeV.

Since in $pA$ collisions no hot and dense medium is expected to be
created, a $pA$ run at the same nucleon-nucleon energy as in
nucleus-nucleus ($AA$) collisions would be of major importance 
to test the theoretical models and to have reliable baseline
spectra for the extraction of novel physical effects. 
A further advantage of $pA$ collisions, is that the multiplicity of
particles in the final state is relatively small compared to $AA$
collisions. For this reason at ALICE minijets may be observed at 
transverse momenta larger than 5 GeV (see Section on ``The experimental
parameters for pA at the LHC" in Ref.~\cite{YelRep}). 
As I will discuss in Sec.~\ref{sec:predictions}, 
this may allow the use of the Cronin effect on minijet production as a
further check of the models.  


\section{THE MODELS}

Soon after the discovery of the Cronin effect \cite{Cronin75}, it was
realized that the observed nuclear enhancement of the $p_t$-spectra
could be explained in terms of multiple interactions
\cite{hardmodel,LP83}. The models may be classified
according to the physical object which is undergoing rescatterings
(the projectile hadron or its partons), and to the ``hardness'' of the
rescattering processes taken into account. Note that a parton is
commonly said to undergo a hard scattering if the 
exchanged momentum is of the order or greater than approximately 1
GeV. However, physically there is no sharp distinction between soft
and hard momentum transfer. Therefore, I prefer to make reference to
the so-called two component models of hadron transverse spectra, and
call hard a scattering which is described by a power-law differential
cross-section at large $p_t$, and soft a scattering whose
cross-section is decreasing faster than any inverse power of the
transverse momentum at large $p_t$. 
In Tables~\ref{tab:hadronic} and \ref{tab:partonic} I provide 
a quick comparison of the hadronic and
partonic rescattering models, respectively.

\subsection{Soft hadronic rescattering models \cite{Wang00} \cite{ZFPBL02} }

These models are based on the pQCD
collinearly factorized cross-section for inclusive particle
production in $pp$ collisions. In order to describe the large-$p_T$
tail of transverse momentum spectra, one has to include also an
intrinsic transverse momentum for the colliding parton. The collision
of a proton on a nuclear target of atomic number $A$ is then obtained
in a Glauber-type model as follows\footnote{Integrations are schematically
indicated with crossed circle symbols. For details see the original
references.}:
\begin{equation}
    \frac{d\sigma^h_{pA}}{d^2p_T} = K \sum\limits_{i,j,k,l}
        F_{i/p} \otimes F_{j/A} 
        \otimes \frac{d\hat\sigma}{d\hat t}(ij\to kl) \, \otimes D^h_k
        \ ,
 \label{pQCD}
\end{equation}
where the proton and nucleus parton distribution functions are,
respectively, 
\begin{align} 
    F_{i/p} = f_{i/p}(x_i,Q^2) \, 
        \frac{e^{-k_{iT}^2/\langle{k_T^2}\rangle_{p}(b)}}
          {\pi\langle{k_T^2}\rangle_{pA}}  
        \hspace*{.4cm}{\rm{and}}\hspace*{.4cm}
    F_{j/A} = T_A(b) \, f_{j/p}(x_j,Q^2) \, 
        \frac{e^{-k_{jT}^2/\langle{k_T^2}\rangle_{Ap}(b)}}
             {\pi\langle{k_T^2}\rangle_{A}}  \ .
 \label{Fsoft}
\end{align}
In Eq.~\eqref{pQCD} $d\hat\sigma/d\hat t(ij\rightarrow kl)$ is
the pQCD parton-parton cross-section, the variables with a hat are
the Mandelstam variables at parton level, and a sum over incoming and outgoing
parton flavours is performed.
The proton is considered point-like compared to the target nucleus,
and to scatter on it at impact parameter $b$. The nucleus is described
by the Woods-Saxon nuclear thickness function $T_A(b)$.
In Eq.~\eqref{Fsoft} $f_{i/p(A)}(x,Q^2)$ are the parton distribution
functions of the proton (nucleus); isospin imbalance is taken into
account and nuclear shadowing is included by the HIJING 
parametrization \cite{hijing}. Partons are assumed to have
an intrinsic transverse momentum with average squared value
$\langle{k_T^2}\rangle_{pA(Ap)}$ and a 
Gaussian distribution. Due to the $k_{i}$ and
$k_j$ integrations a regulator mass $m$$=$$0.8$ GeV has been used
in the pQCD cross-section. Finally, $D_k^h(z,{Q'}^2)$ are the fragmentation
functions of a parton $k$ into a hadron $h$ with a fraction $z$ of the
parton momentum.

Soft proton-nucleon interactions are assumed to excite the projectile 
proton's wavefunction, so that when the proton interacts with the next
target nucleon its partons have a broadened intrinsic 
momentum. Each rescattering of the proton is assumed to contribute to
the intrinsic momentum broadening in the same way, so that: 
\begin{equation}
    \langle{k_T^2}\rangle_{pA}(b,\sqrt s) 
        = \langle{k_T^2}\rangle_{pp} + \delta \times h_A(b,\sqrt s) \ ,
 \label{ktbroad}
\end{equation}
where $\langle{k_T^2}\rangle_{pp}$ is the proton intrinsic momentum needed to
describe hadron transverse spectra in $pp$ collisions, 
$\delta$ is the average momentum squared acquired in each
rescattering, and 
\begin{equation}
    h_A(b,\sqrt s) = \left\{ \begin{array}{ll}
        \displaystyle \nu_A(b,\sqrt s)-1 
            & {\rm\ if\ \ } \displaystyle \nu_A-1 \leq n \\
        \displaystyle n 
            & {\rm\ if\ \ } \displaystyle \nu_A-1 > n 
        \end{array} \right.
 \label{ha}
\end{equation}
represent the average number of collisions which
are effective in broadening the intrinsic momentum. Both models assume
$h_A$ to be a function of the number of proton-nucleon collisions
$\nu_a(b,\sqrt s) = \sigma_{pp}(\sqrt s) T_A(b)$, with $\sigma_{pp}$
the nucleon-nucleon inelastic cross-section. However, Ref.~\cite{Wang00} 
takes $m$$=$$\infty$, while Ref.~\cite{ZFPBL02} assumes an upper limit
$n$$=$$4$ justified in terms of a proton dissociation mechanism: after
a given number of interactions the proton is so excited that it can no
more interact as a whole with th enext nucleon. 
I will call the first model simply {\it soft} and the
second {\it soft-saturated}. 
In both models target nucleon do not have rescatterings, so that 
\begin{equation*}
    \langle{k_T^2}\rangle_{Ap}(b,\sqrt s) 
        = \langle{k_T^2}\rangle_{pp}
\end{equation*}
Further differences between the models are
related to the choices of the hard-scales, of the $K$-factor which
simulates NLO contributions to the parton cross-section, and to the
parametrizations of $\langle{k^2_T}\rangle_{pp}$ and $\delta$ 
(see Table~\ref{tab:hadronic}). 

\begin{figure}[tbp]
\begin{center}
Table~\ref{tab:hadronic} -- 
Parameters of the soft hadronic rescattering models.
\cite{ZFPBL02}.
\begin{footnotesize}
\vskip0.2cm
\begin{tabular}{||l|c|c|c|c|c|c|c||}
\hhline{|t:========:t|}
 \ model 
        & \!\!hard scales {\scriptsize GeV}\!\!
        & \!\!\!$K$\!\! 
        & regul. 
        & proton intrinsic $k_T$ {\scriptsize (GeV$^2$)}
        & \!\!n\!\! 
        & average $k_T$-kick {\scriptsize (GeV$^2$)}
        & nPDF
        \\ \hhline{||--------||}
 \!Soft 
        & $Q=Q'=p_T$ 
        & \!\!2\!\!
        & \!\!0.8 {\scriptsize GeV}\!\!
        & \!\!$\langle{k_T^2}\rangle_{pp} = 
                1.2 + \!0.2 \alpha_s(q^2)q^2 \, ^\dagger\!$ \!\!
        & \!\!$\infty$\!\! 
        & \!$\delta(Q) = 0.255$ 
        {\small $\frac{\ln^2(Q/GeV)}{1+\ln^{\ \!}(Q/GeV)}$}\!\!  
        & \!\!\!HIJING\!\!\!
        \\ \hhline{||--------||}
 \!Soft-sat. \!\!\!
        & \!\! $Q\!=\!${\normalsize$\frac{p_T}{2z_c}$}
        $;\,Q'\!=\!\!${\normalsize$\frac{p_T}{2} ^\flat$}\!\!
        & \!\!1\!\!
        & \!\!0.8 {\scriptsize GeV}\!\!
        & $\langle{k_T^2}\rangle_{pp} 
               = F(\sqrt s) \ ^\natural$  
        & \!4\!\! 
        & $\delta = 0.4$ \!
        & \!\!\!HIJING\!\!\!
\\ \hhline{|b:========:b|}
\end{tabular}
\end{footnotesize}
\parbox{15cm}{\footnotesize
$^\dagger$ $q^2 = 2\hat s \hat t \hat u / (\hat s + \hat t 
+ \hat u)^2 $; parametrization chosen to best reproduce $pp$ data. \\
$^\natural$ No explicit parametrization is given. 
Values of $\langle{k_T^2}\rangle_{pp}$ extracted from a
``best fit'' to $pp$ data, see Fig.15 of Ref.~\cite{ZFPBL02}. \\
$^\flat$ These scales used in the computations of Table~\ref{tab:results}.
In Ref.~\cite{ZFPBL02} $Q\!=\!${\normalsize$\frac{p_T}{2}$} and 
$\,Q'\!=\!\!${\normalsize$\frac{p_T}{2z_c}$}.
}
\end{center}
 \label{tab:hadronic}
\vskip-.6cm
\end{figure}

\subsection{Soft partonic rescatterings: the colour dipole model \cite{KNST02}}

In this model the particle production mechanism 
is controlled by the {\it coherence length} $l_c $$=$$ \sqrt s/(m_N k_T)$,
where $m_N$ is the nucleon mass and $k_T$ the transverse momentum of
the parton which fragments in the observed hadron. 
Depending on the value of $l_c$, three different calculational schemes
are considered.
{\it (a)} In fixed target experiments at low energy (e.g., at SPS), 
where $l_c \ll R_A$, 
the projectile's partons interact incoherently with target nucleons 
and high-$p_t$ hadrons are assumed to originate
mainly from projectile's parton which underwent a hard 
interaction and whose transverse momentum was broadened 
by soft parton rescatterings. The parton is then put on-shell by a
single semihard scattering computed in factorized pQCD. 
This scheme I will discuss in detail below.
{\it (b)} At LHC, where the c.m. energy is very large and $l_c \gg
R_A$, the partons interact coherently with the target nucleons and high-$p_T$
hadrons are assumed to originate from radiated gluons; parton
scatterings and gluon radiation are computed in the light-cone dipole
formalism in the target rest frame. 
{\it (c)} At intermediate energies, like at RHIC, an interpolation is made
between the results of the low- and high- energies regimes discussed above.
All the phenomenological parameters needed in this model are fixed in 
reactions different from $pA$ collisions, and in this sense the model
is said to be parameter-free. 

In the short coherence length scheme, pQCD factorization is
assumed to be valid and formula \eqref{pQCD} is used 
with parton masses $m_g $$=$$ 0.8$ GeV and $m_q$$=$$0.2$ GeV for, viz.,  gluons
and quarks. Moreover, 
\begin{align*} 
    F_{i/p} = 
        f_{i/p} \big( x_i+{\textstyle \frac{\Delta E}{x_a E_p}},Q^2 \big) \, 
        \frac{dN_i}{d^2k_{iT}}(x,b)
        \hspace*{.4cm}{\rm and}\hspace*{.4cm}
    F_{j/A} = T_A(b) \, f_{j/p}\big( x_j,Q^2 \big) \,
        \frac{dN_j^{(0)}}{d^2k_{jT}}(x,b) \ .
\end{align*}
Parton rescatterings are computed in terms of the
propagation of a $q\bar q$ pair through the target nucleus, 
and the final parton transverse momentum distibution
$dN_i/d^2k_{iT}$ is written as:
\begin{align}
    \frac{dN_i}{d^2k_{iT}} & = \!
        \int \! d^2r_1d^2 r_2\,e^{i\,\vec k_T\,(\vec r_1 - \vec r_2)} 
        \left[ \frac{\langle{k_0^2}\rangle}{\pi}\,
        e^{-\frac12(r_1^2+r_2^2) \langle{k_0^2}\rangle}
        \right] 
        \left[ e^{-\frac12\,\sigma^N_{\bar qq}(\vec r_1-\vec r_2,x)
        \,T_A(b)} \right] 
        = \frac{dN^{(0)}}{d^2k_{T}}
        + O\big( \sigma^N_{\bar qq}T_A\big) \, .
 \label{dNdkt}
\end{align}
The first bracket in Eq.~\eqref{dNdkt} represents the contribution of the
proton intrinsic momentum. The second bracket 
is the contribution of soft parton 
rescatterings on target nucleons, expressed through the
phenomenological dipole cross-secton: for a quark,
$
    \sigma_{\bar qq}^N(r_T,x)=\sigma_0\,
        \left[ 1-\exp\big(-\frac{1}{4}\,r_T^2\,Q_s^2(x)\big) \right]
$ 
with $Q_s $$=$$ 1\,{\rm GeV}(x/x_0)^{\lambda/2}$ 
(for the value of the parameters see Table~\ref{tab:partonic}); 
for a gluon $\sigma_{\bar gg}^N $$=$$ 9/4 \, \sigma_{\bar qq}^N$ is used. The expansion of 
Eq.~\eqref{dNdkt} to zeroth order in $\sigma_{\bar qq}^N$ 
gives the intrinsic $k_T$ distribution $dN^{(0)}/d^2k_{T}$ 
of the nucleon partons; the
first order term represents the contribution of one-rescattering
processes, and so on. Partons from the target nucleus are assumed not to
undergo rescatterings because of the small size of the projectile. 
Energy loss of the
projectile partons is taken into account by a shift 
of their fractional momentum proportional to the energy of the
radiated gluons, given by the product of the average
mean path length $\Delta L$ and  the energy loss rate $dE/dz$ \cite{14}.
As nuclear shadowing effects are computed theoretically in the dipole
formalism, see Eq.~\eqref{dNdkt}, 
parton distribution functions in the target are modified only to
take into account antishadowing at large $x$ according to the EKS98
parametrization \cite{EKS98}.

By Fourier transforming the dipole cross-section one sees that the
transverse momentum distribution of the single parton-nucleon
scattering is Gaussian, whence the classification of this model among
the ``soft'' ones. However, the single distributions are not just
convoluted obtaining a broadening proportional to the average number of
rescatterings. Indeed, in the second bracket the rescattering
processes have a Poisson probability distribution. As a result, the nuclear
broadening of the intrinsic momentum is smaller than the product of
the average number of rescatterings and the single scattering
broadening; this might give a
dynamical explanation of the assumption used in the Soft-saturated model
\cite{ZFPBL02}, that $n \,{\scriptstyle\lneqq}\, \infty$ in Eq.~\eqref{ha}.

\begin{figure}[tbp]
\begin{center}
\parbox{15cm}{\begin{center}\small
Table~\ref{tab:partonic} -- 
Parmeters of the soft partonic rescattering model (at short-$l_c$)
and of the hard partonic rescattering models.\vskip-.4cm
\end{center}}
\begin{footnotesize}
\vskip0.2cm
\begin{tabular}{||l|c|c|c|c|c|c|c||}
\hhline{|t:========:t|}
 \ model 
        & \!\!hard scales\!\!
        & \!\!$K$\!\!
        & \!\!regulators (GeV)\!\!
        & \!\!intr. $k_T$\!\!
        & \!\! $dE/dz$ \!\!
        & \!\!dipole cross-sect.\!\!
        & \!\!nPDF\!\!
        \\ \hhline{||--------||}
 \!Col. dip. \!\!\! 
        & \!\!$Q=Q'=p_T$\!\! 
        & \!\! \scriptsize$\otimes$ \!\! 
        & \!\!$m_g$=0.8, $m_q$=0.2\!\!
        & \!\!as Soft mod.\!\!
        & \!\!\!\! -2.5\,GeV/fm\!\!
        & \!\!$\sigma_0$=23\,mb, $\lambda$=0.288, $x_0$=3$\cdot 10^{-4}$\!\!
        & \!\!EKS98$^\dagger$\!\! 
        \\ \hhline{||--------||}
 \!Hard AT
        & \!\!$Q=Q'=p_T^*$\!\!
        & \!\! 2 \!\! 
        & \!\!$\mu$ free param.\!\!
        & \!\!no\!\!
        & \!\!no\!\!
        & \!\!computed from pQCD\!\!
        & \!\!no\!\!
        \\ \hhline{||--------||}
 \!Hard GV
        & \!\!$Q=Q'=p_T$\!\! 
        & \!\! \scriptsize$\otimes$ \!\! 
        & \!\!$\mu$=0.42$^\flat$\!\!
        & \!\!\!as Soft mod.\!\!
        & \!\!no\!\!
        & \!\!----\!\!
        & \!\!\!EKS98\!\!\!
\\ \hhline{|b:========:b|}
\end{tabular}
\end{footnotesize}
\parbox{15cm}{\footnotesize
$^\star$ $Q$$=$$\mu$ in Ref.~\cite{AT01}.\ \
$^\dagger$ Only at large $x_j$ (EMC effect).\ \
$^\otimes$ Factors out in the Cronin ratio. \\
$^\flat$ $\mu$ determines only the typical momentum transfer in elastic 
rescatterings.}
\end{center}
 \label{tab:partonic}
\vskip-.6cm
\end{figure}

\subsection{Hard partonic rescattering model \cite{AT01} \cite{GV02}}
\label{sec:hardresc}

The model of Ref.~\cite{AT01}, hereafter labeled ``hard AT'', 
assumes parton rescatterings responsible of the Cronin
enhancement, and includes up to now in the computations only 
semihard scatterings, i.e., scatterings 
described by the pQCD parton-parton cross-section.
It is the generalization to an arbitrary number of hard parton
rescatterings of the early models of 
Refs.~\cite{hardmodel,LP83,Kastella87} and of the more recent 
Refs.~\cite{Wang97,WW01}, limited to 1 hard rescattering. 
As shown in Ref.~\cite{LP83}, considering only one rescattering 
may be a reasonable assumption at low energy to describe 
the gross features of the Cronin effect, but already
at RHIC energies this might not be enough for the computation of the
Cronin peak $R_M$ \cite{AT01}. The AT model 
assumes the S-matrix for a collision of $n$ partons from the
on $m$ partons from the target to be 
factorizable in terms of S-matrices for parton-parton
elastic-scattering, and assumes generalized pQCD factorization
\cite{hightwist}. The result is 
a unitarized cross-section, as discussed in Refs.~\cite{AT01,BFPT02}:
\begin{align}
    \frac{d\sigma^h_{pA}}{d^2p_T} = \sum\limits_{i}
        f_{i|p} \otimes \frac{d N_{i|A}}{d^2k_T} \, \otimes D^h_i
        + \sum\limits_{j}
        f_{j|A} \, T_A \otimes \frac{d N_{j|p}}{d^2k_T} \, \otimes D^h_j
        \ .
 \label{hardresc}
\end{align}
The first term accounts for multiple semihard scatterings of proton partons on
the nucleus; in the second term the nucleus partons are assumed to undergo a
single scattering, and  
$\frac{d N_{j|p}}{d^2k_T}=\sum_i f_{i|p} \otimes \sigma_{i|H}^N$. 
Nuclear effects are included in $dN^H_i/d^2k_T$, 
the average transverse momentum distribution of a 
proton parton who suffered {\it at least} one semihard scattering.
In impact parameter space it reads 
\begin{align}
    \frac{d N_{i|A}}{d^2k_T}(b) = \int \frac{d^2r}{4\pi}
        e^{\,-i \vec{k}_T \cdot \vec r} 
        \left[  e^{\,-\sigma_{i|H}^N(r)T_A(b)} - 
        e^{\,-\sigma_{i|H}^N T_A(b)}  \right] \ ,
 \label{Harddip}
\end{align}
where unitarity is explicitly implemented at the
nuclear level, as discussed in Ref.~\cite{BFPT02}. 
In Eq.~\eqref{Harddip} 
$
    \sigma_{i|H}^N(r) = K \sum_j \int d^2 p 
        \left[ 1 - e^{\,-i \vec p \cdot \vec r} \,\right]
        \frac{d \hat\sigma}{d \hat t}$ 
        {\small \!\!$(ij \rightarrow ij) \otimes f_{j|p}$}. 
Moreover, $\omega_i \equiv \sigma_{i|H}^N T_A(b)$ (i.e., the
parton-nucleon semihard cross-section times the thickness function) is
identified with the target opacity to the parton propagation.
Note that $\sigma_{i|H}^N(r) \propto r^2$ as $r\rightarrow 0$ and
$\sigma_{i|H}^N(r) \rightarrow \sigma_{i|H}^N$ as $r\rightarrow \infty$. This,
together with the similarity of Eqs.~\eqref{Harddip} and \eqref{dNdkt}
suggests the interpretation of
$\sigma_{i|H}^N(r)$ as a {\it hard dipole cross-section}, which
accounts for hard parton rescatterings analogously to what
$\sigma_{q\bar q}^N$ does for soft parton rescatterings.
Note that no nuclear effects on PDF's are included, but
shadowing is partly taken into account by the multiple scattering
processes.

To regularize the IR divergences of the pQCD cross-sections a small mass
regulator $\mu$ is introduced in the parton propagators, and is
considered a free parameter of the model which signals the scale at
which pQCD computations break down. As a consequence of 
the unitarization of the interaction, due to the inclusion of  
rescatterings, 
both $p_\times$ and $p_M$ are almost insensitive on $\mu$
\cite{AT01}. For this reason these two quantites are considered 
a reliable prediction of the model\footnote{This result is
very different from the conclusion of Ref.~\cite{WW01}, based on a
single-rescattering approximation, that $p_\times \propto
p_0$. Hence $p_\times$ cannot be used to ``measure'' the onset of hard
dynamics as proposed in that paper.}. Note, however, 
that they both depend on the
c.m. energy $\sqrt s$ and on the pseudorapidity $\eta$. 
On the other hand, $R_M$ is strongly sensitive to the
IR regulator. This sensivity may be traced back to the inverse-power
dependence on $\mu$ of the target opacity $\omega_i$:
$\omega_i\propto 1/\mu^a$, where the power $a>2$ is energy and
rapidity dependent. The divergence of $\omega_i$ as $\mu\rightarrow 0$
indicates the need of unitarization of the parton-nucleon
cross-section, and deserves further study. Therefore, $\mu$ can be here
considered only an effective scale which simulates non-perturbative
physics not considered here \cite{EKRT00,Iancu02}, the non-linear evolution of
parton distribution functions in the target \cite{EHKRS02}
and physical effects up to now
neglected, e.g., collisional and radiative energy losses \cite{enloss}.

In the model of Ref.~\cite{GV02}, hereafter labeled ``hard GV'', the
transverse momentum broadening of a parton which undergoes semihard
rescatterings is evaluated with the help of Eq.~\eqref{Harddip} to be
\begin{align}
  \langle k_T^2 \rangle_{H} = \omega \mu^2 
    \ln\left(1+c\frac{p_T^2}{\mu^2}\right)
  \ , 
 \label{GVkt2} 
\end{align}
where the IR regulator $\mu$  
is physically identified with the medium
screening mass and fixed to $\mu=0.42$ GeV, and represents the 
typical momentum kick in each elastic rescattering of a hard parton. 
The factor $c$ and the
constant term 1 are introduced in order to obtain no broadening for
$p_T\approx 0$ partons, as required by kinematic considerations.
The average value in the
transverse plane of the opacity $\omega\approx(0.4/\rm{fm})R_A$ (with $R_A$
the nuclear radius) and of the factor $c/\mu^2=0.18$ are fixed in order to
reproduce the experimental data at $\sqrt s$=27.4 GeV 
and $\sqrt s$=38.8 GeV \cite{data400lab}. With these values the logarithmic 
enhancement in Eq.~\eqref{GVkt2} is of order 1 for $p_T\approx3$ GeV.
Note that $\omega$ and $c$ are assumed to be independent of $\sqrt s$
and $\eta$. Finally, the transverse spectrum is computed by using
Eq.~\eqref{pQCD} and adding to the semihard broadening of
Eq.~\eqref{GVkt2} the 
intrinsic momentum of the projectile partons, 
$\langle k_0^2 \rangle$=1.8 GeV$^2$:
\begin{equation*}
  \langle k_T^2 \rangle_{pA} = \langle k_T^2 \rangle_{pp} 
    + \langle k_T^2 \rangle_H \ ,
\end{equation*}
with shadowing and antishadowing corrections to target partons as in
the EKS98 parametrization \cite{EKS98}.


\section{PREDICTIONS AND CONCLUSIONS}
\label{sec:predictions}

In Table~\ref{tab:results} 
I listed the values of the Cronin parameters $p_M$ and
$R_M$ computed in the various models for proton-nucleon center-of-mass energy
of 27.4 GeV (representing the low-energy experiments at CERN ISR and
SPS and at Fermilab), of 200 GeV (RHIC) and of 5500 GeV (LHC). The
targets considered in the Cronin ratio are listed as
well. Since $p_\times$ is of the order of 1 GeV in almost all models
at all energies, and lies at the border of the validity range of
the models, it's value is not shown. 
Uncertainties of the model calculations are included when
discussed in the original references (see the notes at the foot of the
table). 
In the case of the soft-saturated model, the uncertainty due to the choice of
shadowing parametrization is illustrated by giving the results obtained with no
shadowing, beside the results obtained with the HIJING parametrization
\cite{hijing}. 
Using the ``new" HIJING parametrization \cite{newHIJING}
would change mid-rapidity results only at LHC energy, where a 15\% smaller 
Cronin peak would be predicted \cite{LevaiPC}\footnote{Note however 
that the new parametrization,
which predicts a much larger gluon shadowing at
$x\lesssim10^{-2}$ than the ``old" one \cite{hijing},  
seems ruled out by data on the ratio of Sn and Ca $F_2$
structure functions \cite{SalgMoriond}.}.  
In the case of the ``hard AT" rescattering
model, the major theoretical uncertainty lies in the choice of the effective
parameter $\mu$ as discussed at the end of Sec.~\ref{sec:hardresc}.
In the table, two choices are presented: {\it (a)} an
energy-independent value $\mu=1.5$ GeV, which leads to an increasing
Cronin effect as energy increases; 
{\it (b)} $\mu$ is identified with the IR cutoff $p_0$ discussed in 
\cite{EH02}, in the context of a leading order pQCD analysis of $pp$
collisions. That analysis found $p_0$
to be an increasing function of $\sqrt s$. By performing a 
simple logarithmic fit to the values extracted from data in Ref.~\cite{EH02} we
find $\mu = p_0(\sqrt s) = 0.060 + 0.283 \log(\sqrt s)$, leading to a 
decreasing Cronin effect. Note that a scale increasing with $\sqrt s$ appears
naturally also in the so-called ``saturation models" for hadron production in
$AA$ collisions \cite{EKRT00,KN01,Acc01}.
In the ``hard GV" model at LHC energy, the remnants of Cronin effect at $p_T
\sim 3$ GeV are overwhelmed  by shadowing and the calculation in this region
cannot be considered reliable as yet. The $R_M=1.05$ at $p_T\simeq 40$ in
Ref.~\cite{GV02} is 
a result of antishadowing in the EK98 parametrization and is not related
to multiple initial state scatterings.

\begin{figure}[tbp]
\begin{center}
Table~\ref{tab:results} -- 
Cronin effect at $\eta $$=$$ 0$: comparison of theoretical
models. $p_M$ is expressed in GeV.\\
\vskip0.2cm
\begin{small}
\begin{tabular}{||c|l||c|c|c||c|c|c||}
\hhline{|t:========:t|}
 \multirow{2}{2.1cm}{\centerline{$\sqrt s$}} 
        & \multirow{2}{2cm}{\centerline{model}} 
        & \multicolumn{3}{|c||}{charged pions} 
        & \multicolumn{3}{c||}{partons} 
        \\ \hhline{||~|~|---||---||}
 && $p_M$  & $R_M$ & Ref. 
        & $p_M$ & $R_M$ & Ref. 
\\ \hline\hline
 \multirow{5}{2.1cm}{\begin{center}
        27.4 GeV \\
        $p$W$\big/$$p$Be 
        \\data: Ref.~\cite{data400lab}
        \end{center}} 
 & Soft & 4.0  & 1.55$^\star$ \hspace*{0.95cm} & \cite{Wang00} 
        &  &  &  \\ 
 & Soft-saturated & 4.5$^\star$; 4.4$^\odot$ & 1.46$^\star$; 1.46$^\odot$ 
        & \cite{LevaiPC} 
        & 5.1$^\star$; 5.1$^\odot$ & 1.50$^\star$; 1.51$^\odot$ 
        & \cite{LevaiPC}  \\
 & Color dipole & 4.5 & 1.43$\pm$0.08$^\otimes$ & \cite{KNST02} 
        &  &  &  \\
 & Hard AT &  &  & 
   & 6 $\pm$ 0.8\,$^\dagger$ & 1.1$^\flat$;\ 1.4$^\natural$ &  \\ 
 & Hard GV & 4 & 1.4 & \cite{GV02}
   &  &  &   
\\ \hhline{=:=======:} 
 \multirow{5}{2cm}{\begin{center}
        200 GeV $p$Au$\big/$$pp$
        \end{center}} 
 & Soft  & 3.5 & 1.35$\pm$0.2$^\ddagger$ & \cite{Wang00} 
        &  &  & \\ 
 & Soft-saturated & 2.9$^\star$; 2.7$^\odot$ & 1.15$^\star$; 1.47$^\odot$ 
        & \cite{LevaiPC} 
        & 4.4$^\star$; 4.2$^\odot$ & 1.29$^\star$; 1.70$^\odot$ & 
        \cite{LevaiPC}  \\
 & Color dipole & 2.7 & 1.1 & \cite{KNST02} 
        &  &  &  \\
 & Hard AT &  &  & 
        & 7$\pm$1\,$^\dagger$  & 1.25$^\flat$;\ 1.2$^\natural$ 
        &  \\
 & Hard GV & 3.0 & 1.3 & \cite{GV02}
        &  &  &   
\\ \hhline{=:=======:} 
 \multirow{5}{2cm}{\begin{center}
        5500 GeV $p$Pb$\big/$$pp$
        \end{center}} 
 & Soft  & 3.5 & 1.08$\pm$0.02$^\ddagger$ & \cite{WangPC}
        &  &  &  \\ 
 & Soft-saturated & 2.4$^\star$; 2.2$^\odot$ & 0.78$^\star$; 1.36$^\odot$ 
        & \cite{LevaiPC}
        & 4.2$^\star$; 4.2$^\odot$ & 0.91$^\star$; 1.60$^\odot$  
        & \cite{LevaiPC} \\
 & Color dipole & 2.5 & 1.06 & \cite{KNST02} 
        &  &  &  \\
 & Hard AT &  &   &  
        & 11$\pm$1.3\,$^\dagger$  & 2.1$^\flat$;\ 1.2$^\natural$ 
        & \\
 & Hard GV & $\approx$ 40$^{\,\lozenge}$ & 1.05$^{\,\lozenge}$ & \cite{GV02}
        &  &  &   
\\ \hhline{|b:========:b|}
\end{tabular}
\end{small}
\parbox{14.9cm}{\footnotesize
$^\star$ With HIJING shadowing \cite{hijing}.
$^\odot$ Without shadowing.
$^\otimes$ Error estimated by varying dE/dz within error bars
\cite{Kop02}.\\   
$^\ddagger$ Central value with multiple scattering effects only; error
estimated by using different shadowing parametrizations. \\ 
$^\dagger$ Numerical errors mainly. \
$^\flat$ Using $\mu=1.5$ GeV. \ $^\natural$ Using 
$\mu=0.060 + 0.283 \log(\sqrt s)$, see text. \\
$^{\lozenge}$ Completely dominated by EKS98 shadowing and anti-shadowing;
result considered not reliable as yet, see text.
} 
\end{center}
 \label{tab:results}
\vskip-.8cm
\end{figure}

As discussed in the introduction, experimental reconstruction of
minijets at ALICE should be possible in $pA$ collisions for minijet
transverse momenta $p_T\gtrsim 5$ GeV. 
The $p_T$-spectrum of the partons which will hadronize giving the
observed minijet may be obtained by setting $D_i^h $$=$$ \delta(z-1)$ in
Eqs.~\eqref{pQCD} and \eqref{hardresc}. This may be very interesting, 
because pQCD computations suffer
from large uncertainties in the determination of FF's at large $z$,
where they are only loosely constrained by existing data
\cite{ZFL02}. For this reason I listed in Table~\ref{tab:results} 
also the Cronin
parameters for the case of parton production. However, jet
reconstruction efficiency should be accurately evaluated 
to assess the usefulness of this observable. 

Table~\ref{tab:results}  shows that there are large theoretical
uncertainties in the extrapolation of the Cronin effect from lower
energies to LHC energy. A major source of uncertainty for most of the models 
is the size of nuclear shadowing and anti-shadowing at small $x$, 
see Ref.~\cite{SalgMoriond} for a detailed discussion and comparison 
of the existing parametrizations.  
For example the HIJING parametrizations \cite{hijing,newHIJING} 
predict more gluon shadowing than the EKS98 \cite{EKS98} at small
$x\lesssim10^{-2}$. At LHC this is the dominant region at mid-rapidity and
medium-small transverse momenta. On the other hands the HIJING parametrizations
predict less antishadowing than EKS98 at $x\gtrsim10^{-1}$, which is the
dominant region at large enough $p_T$ at all energies. 
At LHC all these effects may lead up to a factor 2 uncertainty in the height of
the Cronin peak $R_M$.

In conclusion, a $pA$ run at LHC is necessary 
both to test theoretical models for particle production 
in a cleaner experimental situation than in $AA$ collisions, 
and to be able to make reliable extrapolations to $AA$ collisions, 
which is the key to
disentangle known effects and new physics. Since, as we have seen, 
the nuclear effects are potentially large, 
it would be even preferable to have a $pA$ run
at the same energy as the $AA$ run. 
In addition, the A systematics, or the study of collision centrality
cuts, would be interesting since would allow to change the opacity of
the target -- then the size of the Cronin effect -- in a controllable way.
Finally, 
let me remark that the $\eta$-systematics of the Cronin effect has been
considered in the literature only in Ref.~\cite{AT01}. 
However, as discussed also in \cite{VitTP}, given the large
pseudorapidity coverage of CMS this observable might 
be a very powerful handle for the understanding of the effect. 
It would allow to systematically scan nuclear targets in the low-$x$ region, 
and would help to test the proposed models, in which the rapidity affects the
size of the Cronin effect in rather different ways.


\vskip.3cm
\noindent

\section*{ACKNOWLEDGEMENTS}
I would like to thank M.~Gyulassy, B.~Kopeliovich, P.~L\'evai, D.~Treleani,
X.~N.~Wang, I.~Vitev and F.~Yuan for many helpful discussions. 
This work is partially 
funded by the European Commission IHP program under contract 
HPRN-CT-2000-00130. 


\end{document}